\documentstyle{aipproc}
\newcommand{\mP}{{\bar M}_P}
\begin{document}

\def\grtsim{\,\,\rlap{\raise 3pt\hbox{$>$}}{\lower 3pt\hbox{$\sim$}}\,\,}
\def\lsim{\,\,\rlap{\raise 3pt\hbox{$<$}}{\lower 3pt\hbox{$\sim$}}\,\,}

\begin{flushright}
SLAC--PUB--8767\\
January 2001
\end{flushright}
\bigskip\bigskip

\title{Probing the Geometry of the Universe at the NLC\footnote{Presented at the 5th International Linear Collider Workshop (LCWS 2000),
Batavia, Illinois, October 24-28, 2000.  Work supported by the Department of Energy, 
contract DE--AC03--76SF00515.}}
\author{Hooman Davoudiasl}
\address{Stanford Linear Accelerator Center, Stanford University, Stanford, 
California 94309, USA}
\maketitle

\begin{abstract}

The hierarchy problem in particle physics has recently been approached from a 
geometric point of view in different models.  
These approaches postulate the existence of extra dimensions with various 
geometric properties, to explain how the hierarchy between the apparent scale of gravity
${\bar M}_P \sim 10^{18}$ GeV and the weak scale $m_W \sim 100$ GeV can be generated.  Generally, 
these models predict that the effects of gravity mediated interactions become strong at
the weak scale.  This fact makes the NLC a promising tool for testing such extra dimensional
models.

\end{abstract}

\section{Introduction}

The gravitational potential $V(r)$ of a test mass $m_t$ at a distance $r$ is observed to be

\begin{equation}
V(r) = - \, G_N \frac{m_t}{r},
\label{V}
\end{equation}  
where $G_N$ is the 4-$d$ Newton's constant.  Thus, gravitational interactions
can be described by a non-renormalizable field theory, where the spin-2 graviton mediates the
force, and couples to the energy momentum tensor with dimensionful coupling $1/\mP$, where 
$\mP \sim G_N^{-1/2}\sim 10^{18}$ GeV.  However, the electroweak interactions have a typical
scale of order the $W$ mass $m_W \sim 100$ GeV.  If the Higgs boson of the Standard Model (SM)
is responsible for the electroweak symmetry breaking, then we expect that the mass of the Higgs
$m_H \sim m_W$.  Thus, $m_H$ seems to be stable against $O(\mP)$ quantum corrections.  Explaining the 
origin of the large ratio $\mP/m_H\sim 10^{16}$ is referred to 
as the hierarchy problem in particle physics.

There have been a number of proposals for solving the hierarchy problem.  An interesting and 
theoretically appealing proposal is low energy supersymmetry.  In a supersymmetric solution,
new fields are added to the Lagrangian, such that every known field will have a superpartner of
weak scale mass.  However, there is, as yet, no experimental evidence for this and other ideas 
requiring the discovery of new particles around the weak scale.

A new class of ideas approaches the hierarchy problem from a geometric point of view.  Instead of 
postulating extra fields, such as in supersymmetry, one postulates the existence of extra
dimensions in the universe.  Here, we present two models that approach the question of 
hierarchy from an extra dimensional viewpoint.  The first one, due to Arkani-Hamed, Dimopoulos,
and Dvali (ADD) \cite{ADD} uses $n > 1$ large extra dimensions; we only briefly discuss this model.  
The second model, due to 
Randall and Sundrum(RS) \cite{RS}, assumes a warped 5-$d$ universe, and is the main subject of what 
follows.

\section{Large Extra Dimensions}

In the ADD model, the assumption is that the fundamental scale of gravity in $(4 + n)$-$d$ is 
$M_F$.  The gravitational potential $V(r)$ at distances $r \gg R$, 
where $R$ is the typical size of the extra dimensions, is given by Gauss' law 
\begin{equation}
V(r) = - \, G_N \frac{m_t}{M_F^{(2 + n)} R^n \, r}.
\label{V4+n}
\end{equation}
To recover the observed gravitational force, we must have 
\begin{equation}
\mP^2 \sim M_F^{(2 + n)} R^n.    
\label{mp}
\end{equation}  
Now, if we require that the $M_F \sim m_W$, in order to eliminate the hierarchy between the two scales, 
we are forced to have large extra dimensions of size 1 fm$\lsim R \lsim$ 1mm , for 
$2 \leq n \leq 6$.  The case $n = 1$ is ruled out, since it requires $R \sim 1$ AU, which would result in 
deviations in Newtonian gravity at the scale of the solar system.

In the ADD scenario with large extra dimensions, (i) there is a Kaluza-Klein tower of gravitons with mass
$m_n \sim n/R$ with equal spacing; (ii) each KK mode couples with $1/\mP$ in 4-$d$; (iii) the KK tower at
energies $\sqrt s$ $\sim M_F \sim 1$ TeV interacts strongly, only suppressed by $1/M_F$, due to the KK
multiplicity of $O(10^{16})$; (iv) the SM resides on a 4-$d$ wall in a $(4 + n)$-$d$ bulk; (v) the 
geometry is factorizable, and the $n$ extra dimensions are flat, that is the metric is of the form 
\begin{equation}
ds^2 = \eta_{\mu \nu} \, dx^\mu dx^\nu + \sum_{i = 4}^{3 + n} dx_i^2.
\label{addmetric}
\end{equation}

\section{The RS Model}

This model \cite{RS} is based on a 5-$d$ spacetime of constant negative curvature, called $AdS_5$, 
truncated by two 4-$d$ Minkowski walls, separated by a fixed distance $L = \pi \, r_c$ 
with $r_c \sim \mP^{-1}$ as the compactification scale; the $5^{th}$
dimension $y$ is parameterized by an angular variable $\phi \in [-\pi, \pi]$ 
and $y = \phi \, r_c$.  The geometry is required to respect the $Z_2$ symmetry $\phi \to -\phi$.
The ``Planck wall'' is at 
$\phi = 0$, whereas the ``SM wall'', corresponding to the visible 4-$d$ universe,  is at $\phi = \pi$.  
The energy density on the Planck wall $V_P$ is equal and opposite to that on the SM wall and we have 
$V_P \sim M_5^3 k$, where $M_5 \sim \mP$ is the fundamental 5-$d$ scale, and $k \sim \mP$ is 
the curvature scale.  The 5-$d$ cosmological constant is given by $\Lambda_5 = - k V_P$.  
Thus, we see that the parameters of the model do not establish new hierarchies.

The geometry of this model is warped and non-factorizable, with the metric
\begin{equation}
ds^2 = e^{-2 \sigma(\phi)} \eta_{\mu \nu} \, dx^\mu dx^\nu + r_c^2 \, d\phi^2 \, \, ; \, \, 
\sigma(\phi) = k \, r_c \, |\phi|, 
\label{rsmetric}
\end{equation} 
where $e^{-2 \sigma(\phi)}$ is the warp factor.  This geometric warp factor offers a possible explanation of 
the hierarchy problem.  Basically, if one writes down a 5-$d$ action with Planckian mass parameters  
$m_5 \sim \mP$, after a KK reduction to 4 dimensions, the 4-$d$ fields with canonical 4-$d$ kinetic 
terms will have mass parameters $m_4 = m_5 \, e^{- k \, r_c \pi}$.  To have $m_4 \sim m_W$, 
we only need to require $k r_c \sim 10$, which has been shown to be easily realized
in a mechanism that stabilizes the size of the $5^{th}$ dimension \cite{GW2}.  In this way, numbers 
of $O(10)$ generate large hierarchies of $O(10^{16})$ 

This model has features that are quite distinct from the ones of the ADD model.  
In the RS model \cite{RS,DHR1} 
(i) the KK tower of gravitons starts at $m \sim 1$ TeV, the spacings between the tower masses 
$\Delta m \sim 1$ TeV are unequal and given by roots of Bessel functions; (ii) the zero mode (massless 
4-$d$) graviton couples with $1/\mP$ and the massive KK tower gravitons couple with $1/\Lambda_\pi 
\sim 1$ TeV$^{-1}$; this can be understood by noting that the wavefunction of the zero mode 
along the $5^{th}$ dimension is localized near the Planck wall, characterized by $\mP$, whereas the KK 
graviton wavefunctions are localized near the SM wall, characterized by $\Lambda_\pi$.  
The RS-type models are sometimes referred to as ``Localized Gravity'' models.  (iii) In the original 
proposal by Randall and Sundrum \cite{RS}, the SM fields are taken to reside only on the SM wall.  
With these features, the RS model predicts resonant production of KK gravitons 
at ${\sqrt s} \sim 1$ TeV at colliders such as the NLC.

There have been a number of generalizations and extensions of the original RS proposal.  Some of these
extensions study the possibility of having SM fields in the bulk and deriving the 4-$d$ physics from the
5-$d$ picture \cite{GW1}, since the SM scale of order 1 TeV can be generated on the SM wall through the
warp factor.  For various phenomenological reasons it is least problematic to keep the Higgs field on the
SM wall \cite{RSf,GP,DHR3}.  However, as a first step, one can study the effect of placing the SM gauge
fields in the bulk and keeping the fermions on the SM wall \cite {DHR2,P}.  In this case, one finds that the
fermions on the wall couple to the KK gauge fields $\sqrt {2 k r_c \pi} \sim 10$ times more strongly than they
couple to the zero mode gauge fields ($\gamma, g, W^{\pm}, Z$).  Here one expects to get strong
constraints from data, and indeed  agreement with precision electroweak data requires that the lightest KK
gauge boson have a mass $m_1^{(A)} \grtsim 23$ TeV.  This value pushes the scale on the SM wall far
above $\sim 1$ TeV, making this scenario disfavored in the context of the hierarchy problem.

The above bound can be somewhat relaxed, if the fermions also reside in the bulk \cite{RSf}.  In fact, by 
introducing bulk fermion 5-$d$ masses $m_\Psi$, one can change the couplings of the fermion zero 
modes (observed SM fermions) to various KK fields, and place different bounds on the RS model, depending 
on the  value of the bulk mass parameter $\nu \equiv m_\Psi/k$ \cite{GN,GP,DHR3}.  The parameter $\nu$ 
controls the shape of the fermion zero mode wavefunction $f^{(0)} \sim e^{\nu \sigma(\phi)}$.  Thus, 
negative values of $\nu$ localize $f^{(0)}$ near the Planck wall, whereas positive values 
of $\nu$ localize $f^{(0)}$ near the SM wall. 

Avoiding large FCNC's may require keeping the value of $\nu$ nearly universal for all fermions.  
The 4-$d$ Yukawa couplings of SM fermions depend on the value of
$\nu$.  It can be shown that keeping the 5-$d$ Yukawa couplings $\lambda_5 \sim 1$ requires that the $\nu
\grtsim - 0.8$ (for the lightest fermion) \cite{GP,DHR3}.  Avoiding the generation of a new hierarchy, on the
other hand, forces us to have $\nu \lsim - 0.3$.  This can be understood by noting that as $\nu$ increases,
the fermions get more and more localized toward the SM wall, and this takes us back to the case where
leaving the fermions on the wall gave us a large lower bound on the mass of the first KK gauge field
\cite{DHR3}. Although placing the SM gauge and fermion fields in the 5-$d$ bulk results in a rich
phenomenology \cite{DHR3}, the range where the theory seems viable is rather narrow, and keeping the SM
fields on the wall appears to be less problematic.

\section{Conclusions}

The hierarchy problem can be approached from a geometric $(4 + n)$-$d$ point of view, in the ADD and 
the RS scenarios.  Each scenario predicts a distinct set of signatures at $\sqrt s \sim 1$ TeV.  
Thus, an NLC with $\sqrt s \sim 1$ TeV can test these ideas and possibly yield information on the 
geometry of the extra dimensions of the universe by probing such features as their number, size, 
and curvature.

\end{document}